\documentclass{appolb}
\usepackage{amsfonts}
\usepackage{amssymb}
\usepackage{amsmath}
\usepackage{cite}
\usepackage{multirow}
\usepackage{enumitem}
 
\setlist{nosep}
\usepackage{graphicx}
\newcommand{\bM}{{\boldsymbol M}}
\newcommand{\blambda}{{\boldsymbol \lambda}}
\newcommand{\bv}{{\boldsymbol v}}

\newcommand{\bra}[1]{\langle{#1}|}
\newcommand{\braket}[1]{\langle{#1}\rangle}
\newcommand{\ket}[1]{|{#1}\rangle}

\newcommand{\nn}{\nonumber}
\newcommand{\crn}{\nn \\}

\newcommand{\power}[1]{\times 10^{#1}}

\newcommand{\MPl}{M_\mathrm{Planck}}

\newcommand{\LPl}{\Lambda_{\rm Pl}}

\renewcommand{\MPl}{M_{\rm Pl}}
\newcommand{\mpl}{M_{\rm Pl}}
\newcommand{\tpl}{t_{\rm Pl}}

\newcommand{\gv}{\mbox{GeV}}

\newcommand{\D}{\mathrm{d}}

\newcommand{\wz}{\sqrt{2}}

\newcommand{\mbo}[1]{$#1$}
\newcommand{\semis}{\;;\;\;}

\newcommand{\MSb}{$\overline{\mathrm{MS}}$ }
\newcommand{\ba}{\begin{eqnarray*}}
\newcommand{\ea}{\end{eqnarray*}}
\newcommand{\bea}{\begin{eqnarray}}
\newcommand{\eea}{\end{eqnarray}}
\newcommand{\bary}{\begin{array}}
\newcommand{\eary}{\end{array}}

\newcommand{\sign}{\mbox{sign}}
\newcommand{\epo}{\,.}

\newcommand{\bit}{\begin{itemize}}
\newcommand{\eit}{\end{itemize}}
\newcommand{\cO}{\mathcal{O}}
\newcommand{\cG}{\mathcal{G}}

\newcommand{\mytexttilde}{{\raise.17ex\hbox{$\scriptstyle\mathtt{\sim}$}}}
\newcommand{\cosW}{\cos^2 \Theta_W}

\newcommand{\SU}{\mathit{SU}}

\newcommand{\mz}{M_Z^2}
\newcommand{\mw}{M_W^2}

\renewcommand{\mbo}[1]{$#1$ }
\newcommand{\lpl}{\LPl}

\newcommand{\ha}{\frac12}
\newcommand{\npb}{Nucl.\ Phys.\ B }

\newcommand{\pl}{Phys.\ Lett.\ }
\newcommand{\prd}{Phys.\ Rev.\ D }
\newcommand{\prl}{Phys.\ Rev.\ Lett.\ }

\begin{document}
\title{%
\vskip-3cm{\baselineskip14pt
\centerline{\small DESY 21-073,~~HU-EP-21/12\hfill May 2021}
}
\vskip1.5cm
The Standard Model of particle physics as a conspiracy theory and the possible role of the Higgs boson in the evolution of the early universe
\thanks{To the memory of our distinguished colleague and guide Prof.~Martinus Veltman
  with his remarkable record in particle physics}}
\author{Fred Jegerlehner
\address{Humboldt-Universit\"at zu Berlin, Institut f\"ur Physik,
       Newtonstrasse 15,\\ D-12489 Berlin, Germany\\
Deutsches Elektronen-Synchrotron (DESY), Platanenallee 6,\\ D-15738 Zeuthen, Germany}
}
\maketitle
\begin{abstract}
 I am considering Veltman's ``The Infrared - Ultraviolet Connection''
 addressing the issue of quadratic divergences and the related huge
 radiative correction predicted by the electroweak Standard Model (SM)
 in the relationship between the bare and the renormalized theory,
 commonly called ``the hierarchy problem'' which usually is claimed
 that this has to be cured. After the discovery of the Higgs particle
 at CERN, which essentially completed the SM, an amazing interrelation
 of the leading interaction strengths of the gauge bosons, the
 top-quark, and the Higgs boson showed up amounting that the SM allows
 for a perturbative extrapolation of the running couplings up to the
 Planck scale. The central question concerns the stability of the
 electroweak vacuum, which requires that the running Higgs
 self-coupling stays positive. Although several evaluations seem to
 favor the meta-stability within the experimental and theoretical
 parameter-uncertainties, one should not exclude the possibility that
 other experiments and improved matching conditions will be able to
 establish the absolute stability of the SM vacuum in the future. I
 will discuss the stable vacuum scenario and its impact on early
 cosmology, revealing the Higgs boson as the inflaton. It turns out
 that the Standard Model's presumed ``hierarchy problem'' and
 similarly the ``cosmological constant problem'' resolve themselves
 when we understand the SM as a low energy effective tail that is
 emergent from a cutoff-medium at the Planck scale. ``The Infrared -
 Ultraviolet Connection'' conveyed by the Higgs boson mass
 renormalization appears in a new light when the energy dependence of
 the SM couplings is taken into account. The bare Higgs boson mass
 square then changes sign below the Planck scale where it is
 activating the Higgs mechanism. At the same time, it reveals that
 the SM towards the Planck scale is in the symmetric phase, where the
 Higgs potential provides a high dark energy density triggering
 inflation, and four heavy Higgs bosons which decay and thereby are
 reheating the inflated early universe.
\end{abstract}
\PACS{11.15.-q,11.10.Hi,12.15.Lk,98.80.Cq}

\maketitle
\section{Introduction and Overview}
I review some points advanced by Martinus
Veltman~\cite{Veltman:2008zz,Veltbrighton} which allow us to
understand better what is behind the electroweak (EW) Standard Model's
peculiar structure and how this fits perfectly into attempts to
consider the Standard Model (SM) as a low energy effective theory of a
cutoff fine-grained ``ether'' at the Planck scale $\LPl$. The discovery of
the Higgs boson~\cite{Englert:1964et,Higgs:1964ia} by the ATLAS and
CMS Collaborations at the LHC at CERN~\cite{ATLAS,CMS} opened a new
book in particle physics. We now see the SM in a new light, not so
much because the Higgs boson has been found finally, but much more
because of the very special mass, the Higgs boson proved to have. In
fact the heaviest SM states, the weak gauge bosons, the top-quark, and
the Higgs boson, interact with each other in such a way that the
leading SM couplings fall into a very narrow window which allows for a
stable Higgs vacuum up to the Planck scale and, equally surprising, the SM
remains accessible to perturbation theory up to the Planck scale.

The reason seems to be a kind of self-organized conspiracy between SM
couplings with an amazing balance between bosonic and fermionic
contributions in the renormalization group evolution of parameters as
a function of the energy. With the other couplings given, the Higgs
boson self-coupling $\lambda$ appears to be self-tuned to let us
understand the SM as emergent from the Planck medium. If the Higgs
vacuum remains stable up to the Planck scale, the SM would shape the
dynamics of the early universe, with inflation, reheating, and what
else triggered by the Higgs system. The Higgs potential inevitably
would provide a huge strongly time-dependent cosmological constant
that shapes inflation and consequently also modifies the
time-dependence of the Hubble constant (see
e.g.~\cite{Poulin:2018cxd}). While effective SM parameters change
little after the time of Big-Bang nucleosynthesis, early cosmology may
be shaped in a previously unexpected manner. Indeed, the Higgs boson
is the only SM particle that directly couples to gravity, as a leading
contribution to the energy-momentum tensor in Einstein's field
equations. The relevant object is the Higgs potential
$$V(\phi)=\frac{m^2(\mu^2)}{2}\:\phi^2+\frac{\lambda(\mu^2)}{24}\:\phi^4\,,$$
depending on the renormalization scale $\mu$, and, provided $\lambda$
stays positive, the SM vacuum remains stable.  The Higgs boson mass
renormalization relation $$ m^2_{H0}-m^2_{H} =\delta
m_H^2=\frac{\lpl^2}{(16\pi^2)}\,C(\mu)\,,$$ relating the bare mass
$m_{H0}$ to the renormalized $m_H$, which Veltman addressed
in~\cite{Veltman:1980mj} as ``Infrared - Ultraviolet Connection'',
obtains a new deeper meaning if we are taking into account that SM
parameters are scale-depen\-dent. It happens that the renormalization
counterterm, depending sensitively on the precise actual SM
parameters, shows a zero $C(\mu_0)=0$ at about $\mu_0\sim 10^{16}~\gv$
at which bare and renormalized Higgs masses are coinciding. Below
the zero, where the bare mass square is negative, the SM falls into
the broken phase where the Higgs boson, like all other massive SM
particles, acquires a mass set by the Higgs boson vacuum expectation
value (VEV) $v\sim 246~\gv$ as the relevant scale. Above the zero, in
the symmetric phase, the effective mass is
\mbo{m_0^2=m^2_{H0}/2=(m_H^2+\delta m_H^2)/2} effectively determined
by the dominating counterterm: $$m_0^2\sim \delta m^2 \simeq
\frac{\LPl^2}{32\pi^2}\,C(\mu)\epo$$ Together with the vacuum
expectation value of the potential the Higgs system provides a
sufficient amount of dark energy to trigger inflation, while the four
heavy unstable Higgs states, which are existing in the symmetric
phase, via their decays are reheating the universe. SM reheating is
efficient again by peculiarities of SM physics. The existence of the
third family with a strong coupling to the Higgs field let the Higgs
bosons decay predominantly into top/anti-top, top/anti-bottom,
etc. quark pairs and their charge conjugates; this heavy quark
radiation later materializes during the EW phase transition by
cascading into the much lighter quark species which are forming
ordinary matter. While dark energy is an SM product the origin of the
dark matter still remains a mystery in this scenario.

\section{Veltman's ``derivation'' of the Standard Model in the light of the Standard
  Model as a low energy effective field theory}
\label{sec1}
One of Martinus Veltman's dedications has been his drive to find a
renormalizable theory for the weak interactions and to scrutinize the
electroweak theory which after all culminated in the SM of particle
physics. The younger particle physicists learning the SM from
textbooks may have little knowledge about the obstacles one had to
surmount before the established structure had been unveiled as the
true theory of electroweak phenomena. The maybe strange-looking
structure of the SM as a gauge theory based on the local symmetry
group $\cG_{\rm SM}=SU(3)_c\otimes SU(2)_L\otimes U(1)_Y$ (gauge
couplings $g_3$, $g_2$ and $g_1$) broken down to $SU(3)_c \otimes
U(1)_{\rm e.m.}$ by the Higgs mechanism exhibits a pattern that really
cries for explanations. Veltman's curiosity has been focused more on
the question of what makes the SM so unique, rather than to question
its validity and to declare its structure unnatural. His cardinal
point in defending the SM and questioning most of the proposed SM
extensions has been that, due to the minimal Higgs structure, the SM
unavoidably predicts a massless photon~\cite{Weinberg67}, which in
most other models have to be imposed as an extra condition, which then
is causing a fine-tuning issue\footnote{\label{photonsusy} The simplest supersymmetric
model accidentally escapes this problem and predicts a zero photon
mass.}~\cite{Veltman:2008zz}. In a remarkable
paper~\cite{Veltbrighton}\footnote{For related considerations
see~\cite{Others}.}  he has given a partial but very enlightening
answer by listing the general conditions which essentially allow us to
derive the SM. Veltman's list of assumptions is the following: \\

\noindent
1) local field theory\\
2) interactions follow from a local gauge principle\\
3) renormalizability\\
4) masses derive from the minimal Higgs system (one physical scalar only)\\
5) the right handed singlet neutrino $\nu _R$, which we know must exist,\\ \hspace*{4mm} does not carry hypercharge.\\

\noindent
The last assumption looks somewhat ad hoc, but we accept it. The
consequences of the assumptions stated above are
remarkable\footnote{The interested reader really should take the time
and have a look at Veltman's surprisingly simple derivation.}  (see
also~\cite{Jegerlehner:1991dq,Jegerlehner:2018zxm}):
\begin{enumerate}[itemsep=1ex, leftmargin=6mm]
\item
breaking $\SU(2)_L$ by a minimal Higgs sector automatically leads to a global
$U(1)_Y$, which can be gauged,
\item
maximal parity violation of the weak interaction $\SU(2)_L$,
\item with $\Theta_W$ the electroweak mixing angle, and $M_W,\: M_Z$
  the masses\footnote{I denote physical masses by capital the related
  \MSb masses by lower case letters.} of the $W$ and $Z$ bosons, the
  tree-level relation $\rho =\mw /(\mz \cosW)=1$, is a parameter
  independent number and hence not subject to renormalization. It
  derives from the accidental global “custodial” $\SU(2)_{\rm cust}$
  symmetry of the Higgs system.
\item
The existence of the strictly massless photon  (one zero-eigenvalue in the spin-1 boson mass-matrix),
\item
parity conservation of QED,
\item
the validity of the Gell-Mann-Nishijima relation $Q=T_3+\frac{Y}{2}$,
\item
fermion family structure  (lepton-quark conspiracy –\\ $U(1)_Y$ anomaly cancellation), 
\item
charge quantization  (if $Y_{\nu R} = 0$ then $Q_i = T_{3i} + Y_i/2$ fixes $Q_i$).
\end{enumerate}
Here we have to add that the requirement of renormalizability also
enforces\\
 \textbf{the existence of a physical neutral spin 0 particle, the
   Higgs boson\footnote{ If not generated by the Higgs mechanism, the
   particle masses affect renormalizability in the $\SU(2)_L$
   weak-interaction sector, which means that it is the $\SU(2)_L$
   which has to be spontaneously broken by the minimal Higgs system.}.}
\noindent
We do not know why right-handed neutrinos are sterile i.e. do not
couple to gauge bosons; they only couple to the Higgs
field. Nevertheless, this property fits with a minimality principle
``not more than necessary''.  In the SM of electroweak interactions,
neutrinos originally were assumed to be massless although it has not
been required by the SM gauge symmetry structure but had to be imposed
by an extra global $U(1)_e \otimes U(1)_\mu \otimes U(1)_\tau$,
implying that right-handed neutrinos did not exist (or did not couple
to anything) such that neutrinos had to be massless. This is
definitely ruled out by the observation of neutrino oscillations,
meaning that each of the subgroups $U(1)_\ell$ is only partially
conserved. Indeed, lepton-flavor number conservation is not an
emergent property but is resulting from the smallness of the neutrino
masses.  In other words, $e$, $\mu$, and $\tau$ only differ in the way
they couple to the Higgs boson, while they have identical couplings to
the gauge bosons.

Another striking property of the SM is the
GIM-mechanism~\cite{Glashow:1970gm}, the tree-level absence of flavor
changing neutral currents, which in the SM by its renormalizability is
manifestly implemented, but which requires fine-tuning in most
extensions of the SM. The discrete $Z_2$ symmetry (called R-parity)
often imposed on the extended two-doublet Higgs system is not an
emergent property (i.e. renormalizability does not require it).

One of the most important insights into how we perceive nature's
underlying structure has been discovered by Ken Wilson (Nobel Price
1982 for his theory for critical phenomena in connection with phase
transitions) in the early 1970s when he developed the systematic
approach to understanding quantitatively the emergence of
long-distance physics (macroscopic properties) from the short distance
properties of the underlying microscopic systems~\cite{Wilson:1971bg}
(see e.g.~\cite{LausanneLectures1976,Jegerlehner:1976xd}). Quite
generally, the microscopic systems are exhibiting an intrinsic short
distance scale i.e. they represent systems exhibiting a cutoff
$\Lambda$ and allow for a low-energy expansion in
$x=E/\Lambda$. Dropping the suppressed terms (i.e. the positive powers
in x), the expansion yields a low energy effective tail which appears
as an emergent low energy structure.  In case of statistical mechanics
type systems the low-energy effective structure turns out to be given
by a renormalizable Euclidean Quantum Field Theory (QFT) which is
universal for a wide class of systems of different short-distance
properties. A groundbreaking discovery. Thereby, a non-trivial
long-range tail emerges as a $D=4$ world because in $D>4$ only
non-interacting free field tails are left\footnote{In condensed matter
physics this is known as the Landau criterion~\cite{Landau:1937obd},
and is the basis for the $\varepsilon$-expansion about $D=4$ in
critical phenomena~\cite{Wilson:1971dc}.}. The $D-4$ extra dimensions
are hidden, unobservable when watched from far away. No need for
curled-up (compactified) extra-dimensions.  Renormalizable $D=4$
Euclidean QFTs are equivalent to Minkowski QFTs (Osterwalder-Schrader
theorem), the basic structure on which our understanding of particle
physics is relying. Note that this mathematical equivalence is a
property of the long-distance tail and thus a Euclidean $D\geq 4$
dimensional short distance system may naturally show its appearance to
us as a $(3,1)$-dimensional Minkowski space-time, i.e. also
Lorentz-invariance (pseudo-rotations) is an emergent symmetry, means,
also here we got rid of the ``ether''. This could mean that also time
is an emergent property that only exists in the low energy tail
perceived. Thereby renormalizability and analyticity are the key
ingredients. Based on the Euclidean-Minkowskian equivalence Wilson
himself developed the tool for solving non-perturbative hadron physics
based on Euclidean lattice Quantum Chromodynamics (QCD)\footnote{In
contrast to the Planck cutoff regularized SM case (discussed below),
here the cutoff provided by the inverse lattice spacing can't be
chosen very large, which causes serious problems in controlling
lattice artifacts in the continuum limit. Therefore it is mandatory to
keep the $\SU(3)_c$ non-Abelian local gauge symmetry exact on the
lattice, because symmetry violating effects could not be controlled by
an expansion in the cutoff in an efficient manner. In practice, the
expansion parameter $E/\Lambda$ would by far not be small enough such
that the symmetry breaking terms would be suppressed as much that the
symmetry would be restored accurately.}.

Originally, the SM~\cite{Glashow61,Weinberg67,Glashow:1970gm} is a minimal
renormalizable completion of Fermi's weak interaction + QED,
supplemented by QCD~\cite{QCD} and two more fermion families. But now
“what is not capable of surviving at long distances does not exist
there” (Darwin revisited). The SM appears as a natural minimal
emergent structure in a low energy expansion from a cutoff system
sitting at the Planck scale! But this is not the only consequence:
in fact, in this kind of scenario, the SM which we supplemented by the
Planck cutoff, leads to unexpected consequences concerning its
high-energy behavior. The latter is governed by the bare
(unrenormalized) theory, which must have been the relevant theory in
the very hot early universe.

Important points are:
\begin{itemize}[itemsep=1ex, leftmargin=6mm, topsep=0pt]
\item[-] the relationship between the renormalized and bare theory is
  calculable,
\item[-] taking into account the energy dependence of parameters is
  mandatory now, this also concerns possible power corrections
  (relevant operators: Higgs mass term and vacuum energy given by the
  Higgs potential and its VEV $\langle V(\phi)\rangle$,
\item[-] leading irrelevant operators of dim 5 (may affect neutrino
  masses) and dim 6 (baryon number violating, required in
  baryogenesis)~\cite{Weinberg:1979sa} naturally can play significant
  roles during the advent of the EW phase transition expected at about
  $10^{16}~\gv$. Especially the dim 6 operators are completely
  suppressed at experimentally accessible energies, but in case the EW
  phase transition happened at very high energies not too far below
  the Planck mass $\MPl$, they are still significant enough to play
  their role in producing the baryon asymmetry,
\item[-] a surprising outcome of the conspiracies between SM couplings
  is that within the SM charge screening is an exceptional property,
  anti-screening the rule. Apart from the non-perturbative QCD
  confinement regime at energies below the hadron mass scales, the SM
  and in particular, its high energy phase (early cosmology) are under
  control by perturbation theory.
\end{itemize}

I see plenty of evidence that the SM is emergent in the Wilsonian
sense~\cite{Wilson:1971bg} as a renormalizable low energy tail of a
cutoff system, the ``ether'', sitting at the Planck scale $\LPl$.  The
equivalent Planck mass $\mpl=(c\hbar/G_N)^{\frac12}$ $\simeq
1.22\power{19}~\gv$, dimensionally determined by Newtons gravitational
constant $G_N$, is one of the fundamental parameters in physics and
the only one representing a fundamental cutoff equivalent to a minimal
length. In an expansion in $E/\LPl$ most short distance details get
lost and we only see what is not suppressed by positive powers of
$E/\LPl$. Note that the cutoff in our case is not just a tool to
regularize ultra-violet (UV) singularities, but represents a true
physical reality behind what we see from far away.  Such a
UV-completed system allows us to perform a low energy expansion in the
cutoff. Thereby, in the low energy tail the symmetries emerge which
usually are absent within the ``ether''. Since in the low energy
expansion all details concerning the UV behavior are lost, it is not a
problem that details about the UV completion remains unknown. Here the
universality of the long-range tail is a key phenomenon well known
from condensed matter systems. There is a big variety of systems
having an identical tail but differ in the infinite tower of
irrelevant dimension $d_{\cO}>4$ operators. When we attempt to
extrapolate the SM to very high scales we expect the next leading
$d_{\cO}=6$ operators to show up when the expansion parameter
$x=E/\LPl$ is about 0.1 producing a 1\% effect, i.e. SM physics can be
valid up to not far below the Planck scale. When going too close to
the Planck scale the expansion of course ceases to make sense. This
does not mean that we lose control of the physics which derives from
the fine-grained Planck system (think for example of a lattice-type SM
as a prototype; that the symmetries of the continuum SM cannot be
fully kept on a lattice~\cite{Luscher:2000hn} does not harm the
emergent SM as a long-distance tail, as we will argue below).  The SM
together with its cutoff UV-completion we call Low Energy Effective
Standard Model (LEESM)\footnote{While the SM as a renormalizable QFT
by itself makes predictions free from any cutoff effects (for
renormalized and observable quantities parametrized in terms of
measured renormalized parameters), the LEESM is an extension of the SM
where the relationship between renormalized and bare parameters has a
physical meaning, which also has been broached by Veltman's ``The
Infrared - Ultraviolet Connection''~\cite{Veltman:1980mj}.}.

What makes Veltman's ``derivation'' of the SM structure so
instructive?  As mentioned before, the SM gauge and Higgs structures,
and the derived consequences just listed, at first sight, look
unexpected and motivated many possible repairs: mirror fermions, grand
unification, supersymmetric extensions, and many more. However, the
peculiar features of the SM are immediate consequences of the general
assumptions 1) to 5) above, which (besides the last point) are
emergent structures when we accept that the SM is a low energy
effective theory of a far away cutoff system. Maybe the most striking
insight is that non-Abelian local gauge symmetries, mathematics-wise a
self-evident generalization of the Abelian local gauge symmetry
familiar from QED, are emergent. The question about the origin of
non-Abelian local gauge symmetries has attracted attention after 't
Hooft's proof of the renormalizability of gauge
theories~\cite{tHooft71a}. Calculations~\cite{Veltman:1968ki,LlewellynSmith:1973ey,Bell:1973ex,Cornwall:1973tb,Jegerlehner:1994zp,Jegerlehner:1978nk,Jegerlehner:1998kt,Djukanovic:2018pep}
(also see~\cite{Bass:2020gpp}) have shown that non-Abelian symmetry is
a consequence of dropping the non-renormalizable terms showing up in
the high energy behavior of tree-level matrix elements of physical
processes, by imposing a tree-unitarity requirement. In fact in the
LEESM scenario ``dropping'' is an automatic feature that derives from
the strong suppression of the questionable terms. Obviously, the
non-Abelian gauge structures require team-play between entries in
gauge group multiplets. What is more natural than excitation modes of
the hot Planck medium grouped in doublets and triplets as the simplest
choices besides possible singlets?  Still, what at first sight looks
almost impossible, namely that the strange-looking SM structure may
naturally emerge as a long-distance structure turns out to be the
consequence of the fact that in the low energy expansion only a
relatively small number of effective interaction vertices are seen, a
tremendous simplification as we cast away an infinite tower of
power-suppressed terms, related to higher-dimensional operators.  The
renormalizable tail, which we can see, naturally must satisfy all
requirements renormalizability imposes if we, also, accept that this
is to be achieved in the simplest possible (minimal) way. Remarkably,
local non-Abelian gauge symmetry, as well as chiral symmetry, are
mandatory for renormalizability and therefore both symmetries are
naturally emergent in the low energy tail. The crux in this game is
that renormalizability predicts the existence of the Higgs boson!  But
also that the weak interactions are maximally P violating while the
QED sector is strictly P conserving are automatic consequences as a
simple calculation shows~\cite{Veltbrighton}. While the non-Abelian
gauge structure implies the characteristic gauge-cancellations
resulting from the interplay of the coupling strength of all those
different vertices which involve the gauge fields, the emergence of
the important Abelian subgroup $U(1)_Y$ looks much more to be a
mystery. But as the calculations show, also here the minimal Higgs
scenario automatically implies a $U(1)_Y$ symmetry and a massless
photon after the spontaneous $SU(2)_L$ breaking.

There is another hidden symmetry deriving from a minimal Higgs system,
the custodial symmetry, which is at the heart of the tree-level
identity $\rho=\mw /(\mz \cosW)=1$ (up to finite SM radiative corrections
$\Delta
\rho$)~\cite{Veltman:1976rt,Veltman:1994vm,Veltman:1977kh}. The
$\rho$-parameter\footnote{ Here a special feature of the weak
corrections comes into play: the non-decoupling of heavy
particles. This is in contradistinction to the Appelquist-Carazzone
theorem~\cite{Appelquist:1974tg}, which infers that heavy states of
mass $M$ are essentially without a trace at energies sufficiently
below the heavy particle threshold. More precisely, effects are
$O(E/M)$ or smaller, in reactions at energies $E \ll M$. This does not
apply in the weak sector of the SM where masses and couplings are
strongly correlated, while it is valid in QED and QCD. This means for
example that electroweak top-quark contributions do not only show up
above the top-quark threshold. In the approximation $M_t,M_H \gg M_Z$
the top-quark and the Higgs boson give a contribution \bea \rho(0)
&\equiv& G_{\rm
  NC}(0)/G_\mu(0)=1+\frac{3\sqrt{2}G_\mu}{16\pi^2}\,\biggl\{M_t^2 \crn
&&+\left(\frac{M_W^2}{1-M_W^2/M_H^2}\ln
M_H^2/M_W^2-\frac{M_Z^2}{1-M_Z^2/M_H^2}\ln
M_H^2/M_Z^2+\cdots\right)\biggr\}\,,\nn \eea to the $\rho$ parameter
at \textbf{zero} momentum transfer. $G_{\rm NC}(0)$ and $G_\mu(0)=G_\mu$ are
the neutral and charged current effective Fermi-constants at vanishing
momentum, respectively.} is the flagpole for heavy states because
quantum corrections at next to leading order are determined by the
difference of the self-energies of the $Z$ and the $W$ bosons
$$\Delta \rho = \frac{\Pi_Z(0)}{M_Z^2}-\frac{\Pi_W(0)}{M_W^2} +
\mathrm{subleading \ terms}$$ and for dimensional reasons corrections
are expected to be proportional to the mass square $M_X^2$ of the
heavy virtual state of mass $M_X>M_Z$ contributing to the gauge boson
self-energies. Most prominent is the top-quark contribution to $\Delta
\rho$ proportional to $G_\mu M_t^2$~\cite{Veltman:1977kh}. Precision
measurements of $\Delta \rho$ by the LEP experiments provided an
important hint for the discovery of the top-quark at the
Tevatron. Large effects only show up for heavier fermion-doublets with
a large mass splitting like the top-bottom quark doublet ($t$,$b$)
where $m_t\gg m_b$. Since $G_\mu M_f^2=y_f^2/(2\wz)$ where $y_f$ is
the Yukawa coupling of a fermion $f$, $\Delta \rho$ effectively
measures the weak isospin splitting within the ($t$,$b$)-doublet which
results from the difference in the top- and bottom-quark Yukawa
couplings.

The custodial symmetry is responsible also for the absence of leading
virtual Higgs boson effects which could contribute to
$\Delta\rho$. There are no corrections proportional to $G_\mu M_H^2$
but only proportional to the logarithmic term $G_\mu M_W^2\:\log
(M_H^2/M_W^2)$ when $M_H^2 \gg M_W^2$, what made it more difficult to
gather reliable information about the Higgs boson mass from
electroweak precision measurements before the Higgs particle's
discovery. Remains the CP violation, which was discovered as a
per-mille effect in neutral kaon decays and is known as a compelling
condition for the dynamical emergence of the baryon-asymmetry in the
universe during baryogenesis (Sakharov
conditions~\cite{Sakharov:1966aja}). In the SM this is achieved
automatically by the triple-replica family structure which has been
proposed by Kobayashi-Maskawa~\cite{CKM} (see
also~\cite{Glashow:1970gm,KorthalsAltes:1972aq}). The predicted
3-family quark-flavor mixing pattern (CKM -matrix) later has been
confirmed in $B$ meson decays to happen precisely as predicted.

In place of a dogma believing that the misunderstood\footnote{See my
analysis in~\cite{Jegerlehner:2018zxm,Jegerlehner:2013nna}.}
hierarchy-problem (see e.g. ~\cite{'tHooft:1979bh}) is an illness of
the SM which must be cured, e.g., by super-symmetrization of the SM as
one possibility, Veltman has stressed many times the fact that within
the minimal SM the zero photon mass is a prediction, not subject to
renormalization, while in most SM extensions (see however
footnote$^{\ref{photonsusy}}$ and~\cite{DiazCruz:1992uw}) the
prediction is lost and has to be imposed as an extra condition, i.e.,
it has to be fine-tuned~\cite{Veltman:2008zz}. Indeed, most extensions
of the SM require non-minimal Higgs sectors: Two Doublet Higgs Models
(TDHM), SUSY extensions like the Minimal Supersymmetric SM (MSSM),
Grand Unified Theories (GUT), left-right symmetric models, the
Peccei-Quinn approach to the strong CP problem, inflaton models base
on an extra scalar field and many more. The masslessness of the photon
is one of the most basic facts of life (no life otherwise), so why
should we make this fundamental property to be something we have to
arrange by hand?

\noindent
In contrast, a minimalist emergent structure makes the SM pretty
unique:
\begin{itemize}[itemsep=1ex, leftmargin=6mm]
\item
I think we now can well understand how various excitations in the hot
chaotic Planck medium can conspire to develop a pattern like the SM as
a low energy effective structure.
\item
Renormalizability as a consequence of the low energy expansion and the
very large gap between the EW and the Planck scales plus certain
minimality (not too little but not too much e.g. only up to symmetry
triplets) determines the SM structure without much freedom.  One could
expect that as a next extended SM structure confined $\SU(4)$ fermion
quartets could provide a bound state spectrum providing dark
matter~\cite{Appelquist:2014jch}, analogous to how QCD is supplying
the bulk of ordinary matter (98\% binding energy).
\item
Minimality is not a new concept in physics as we know e.g. from the
principle of least action or the SM as a minimal renormalizable
extension of QED plus Fermi theory.
\item
The 3 fermion-families are required so that CP violation emerges
naturally, and to open the possibility that baryogenesis can find an
explanation within the SM.  While normal matter requires sufficiently
light quarks, i.e. small Yukawa couplings, the cooperation of
couplings that eventually allows the SM to extend up to the Planck
scale is only possible when Yukawa couplings in the ballpark of the
gauge-couplings and the Higgs self-coupling exist, which is what is
achieved with the third family only. A fourth family\footnote{ Given
the LEP and LHC 4th family mass bounds at least some of the members
would have to be heavier than the top quark and the corresponding
large Yukawa couplings would imply dramatic changes in the
extrapolation of the SM parameter running. Also, the measured bounds
of the $\rho$ parameter essentially rule out having further fermion
doublets with substantial mass splittings.} obviously would spoil the 
feature of the SM that we can understand it as emergent from a Planck
medium.
\item
It is interesting to note that QCD based on $\SU(3)_c$ local
gauge-symmetry requires massless gluons. Massless gauge fields, in
this case, are not only required by renormalizability but are also a
condition for confinement to work. We also note that, in contrast to
the electroweak sector with its 3 fermion-family structure where CP
violation is automatically generated, CP violation in QCD is obtained
only if we add an extra CP violating term to the Lagrangian. Indeed,
this extra CP violation is not something that the low energy expansion
generates necessarily unless it is inherent as a property of the
``ether'' itself. Indeed, CP violation seems to be absent in strong
interactions and axions may not be required to exist. Keep in mind
that $P$ and $CP$ violations are emergent in the electroweak sector
and certainly are absent in the primordial ``ether'' medium, if not
one would expect to find different $P$- and $CP$-violation patterns in
the SM. Interestingly, also symmetry breaking can be an emergent
feature, like maximal $P$ violation out of a $P$ invariant medium
shows.
\end{itemize}
In what follows I review the most striking consequences of a LEESM
scenario which I worked out in some details
in~\cite{Jegerlehner:2013cta,Jegerlehner:2014mua} (for a summary see
also~\cite{Jegerlehner:2014lba}), for the case that SM couplings are
such that an extrapolation up to $\MPl$ is possible. The difference
between stability and meta-stability depends on a difference in the
calculation of the \MSb top-quark Yukawa coupling from the top-quark
mass~\cite{Beneke:2016cbu}. Because of confinement in QCD, the
top-quark as a colored object is not observable by itself but only a
color-screened state of it. Therefore the on-shell top-quark mass
usually taken as an input in the matching conditions is not what the
experiment sees directly. In the end, the \MSb top-quark Yukawa
coupling is not observable and is not unambiguously fixed by
experiment. It rather depends on theory input, which is not fully
under control because the color screening is a non-perturbative issue.

\section{On the running SM couplings}
\label{sec2}
That for a quantum field theory (QFT) coupling constants are not
constant but are renormalization scale-dependent as prescribed by the
renormalization groups (RGs) is known since renormalization of QFTs is
known. Screening (like Abelian gauge couplings) or anti screening
(like non-Abelian gauge couplings) effects let couplings grow or
diminish as a function of the energy scale, respectively. These
effects are well known and for the QED fine structure constant
$\alpha_{\rm QED}(s)$ (screening) and the QCD strong interaction
constant $\alpha_s(s)$ (anti-screening) experimentally well
established, where $s$ is the center of mass energy square at which a
process takes place. Of interest are the gauge couplings $g_1$, $g_2$
and $g_3$ together with the top-quark Yukawa coupling $y_t$ and the
Higgs self-coupling $\lambda$, where the latter two are determined
from the top-quark mass and the Higgs boson mass, respectively, via
the mass-coupling relations. A crucial point here concerns the
matching conditions which are required to calculate the \MSb
parameters from experimentally determined observables like the
physical masses.

\begin{figure}
\centering
\includegraphics[width=0.45\textwidth]{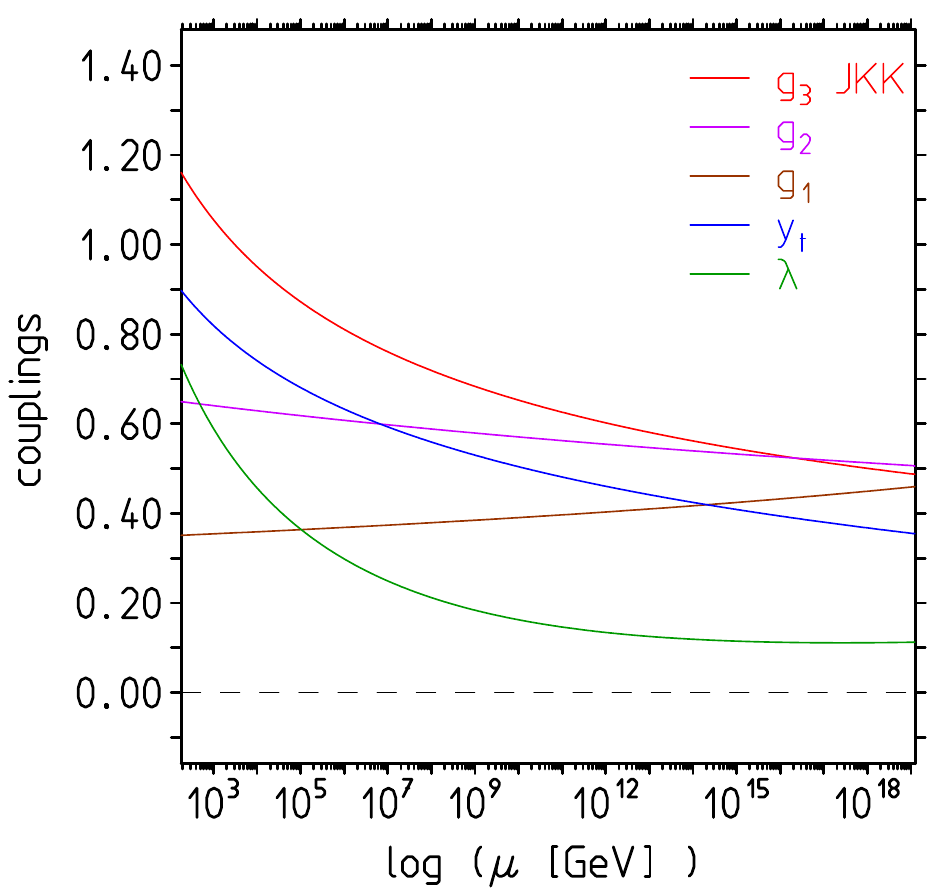}
\includegraphics[width=0.45\textwidth]{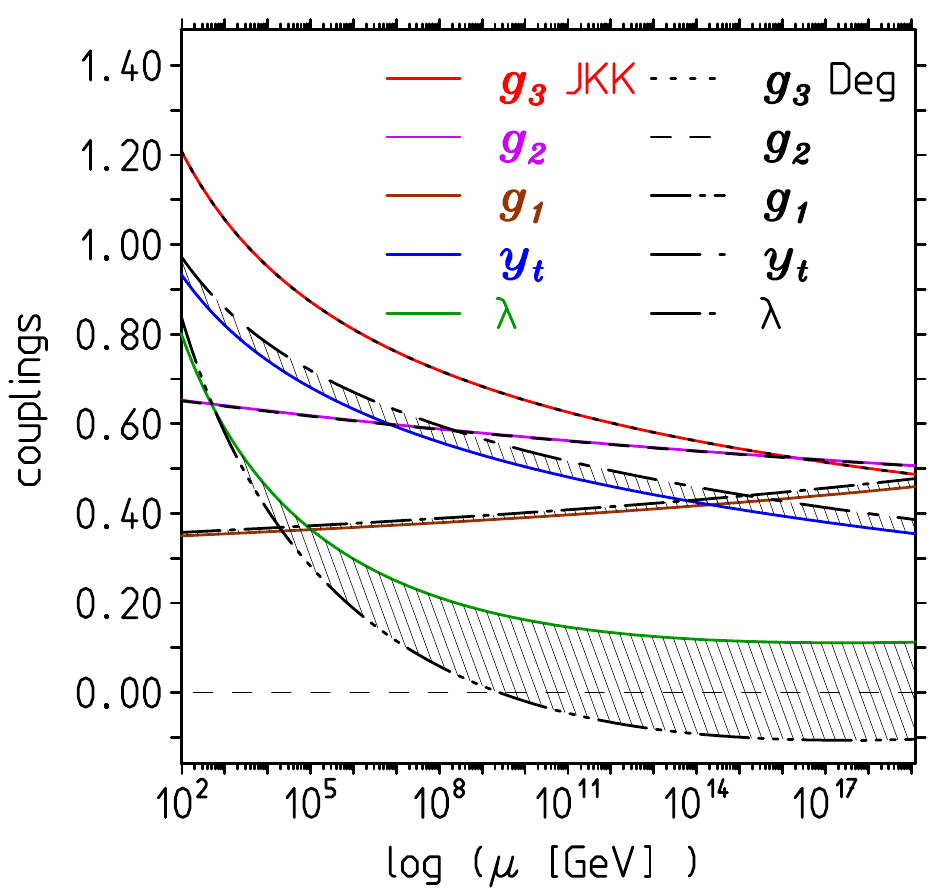}
\caption{ Running SM parameters from the $Z$ mass scale up to
  $\MPl$. Left: on-shell vs \MSb parameter matching based on F.J.,
  Kalmykov, Kniehl~\cite{Jegerlehner:2012kn}(JKK). Right: the same in
  comparison with Shaposnikov et al.~\cite{Yukawa:3}, Degrassi et
  al.~\cite{Degrassi:2012ry} (Deg) matching. Using JKK input
  parameters the vacuum remains stable, with Deg input vacuum
  stability breaks down at about $10^{9}~\gv$ and enters in a
  metastable state of the effective potential~\cite{Coleman:1973jx}
  (i.e. including radiative corrections). Small cause, great impact,
  just by changing a little the input parameters, most sensitive on
  $y_t$, we get a completely different behavior of the Higgs system
  towards the Planck era. The shaded space illustrates the
  parameter-matching controversy, which concerns $y_t$ at $M_Z$. Plots
  adapted from~\cite{Jegerlehner:2013cta}.}
\label{fig:SMrun}
\end{figure}

The discovery of the Higgs boson with a mass of about 125 GeV revealed
an amazing conspiracy between the gauge couplings, the top-quark
Yukawa coupling and the Higgs boson self-coupling. These leading
couplings turn out to be falling into a narrow
window~\cite{Lindner:1988ww,Sher:1988mj,Casas:1994qy,Hambye:1996wb}
which allows us to extrapolate the SM parameters up to the Planck
scale as displayed in Fig.~\ref{fig:SMrun}. Whether this window is
matched perfectly or is almost missing the bottom of the stability
valley is a matter of controversy and depends on the implementation of
the matching conditions which are required to calculate the \MSb
parameters in terms of physical couplings and/or masses.  While most
analyses are based on~\cite{Yukawa:3,Degrassi:2012ry} and predict a
meta stable effective Higgs potential (see
also~\cite{Bednyakov:2015sca,Kniehl:2015nwa,Bass:2020egf}), a slightly
modified evaluation of \MSb parameters based
on~\cite{Jegerlehner:2012kn} revealed vacuum
stability~\cite{Jegerlehner:2013cta,Jegerlehner:2018zxm}. I adopt the
view:\\ \textit{``Although other evaluations of the matching
  conditions seem to favor the meta-stability of the electroweak
  vacuum within the experimental and theoretical uncertainties, one
  should not exclude the possibility that other experiments and
  improved matching conditions will be able to establish the absolute
  stability of the Standard Model in the future.''}\\
Very exciting, and for me it has been completely unexpected, all SM
couplings except the $U(1)_Y$ coupling $g_1$ behave like non-Abelian
ones, i.e., they show anti-screening properties~\cite{AF}. This means
that perturbation theory works and gets better if we extrapolate to
the Planck scale. Taking the \MSb couplings at the $Z$ boson mass
scale as~\cite{Jegerlehner:2013cta,Jegerlehner:2018zxm} $g_1\simeq 0.350$, $g_2\simeq
0.653$, $g_3\simeq1.220$, $y_t\simeq 0.935$ and $\lambda\simeq0.807$,
the following picture arises\footnote{The RG equations $\mu \frac{d}{d
  \mu}\: g_i(\mu)=\beta_i(g_i)\;(i=1,2,3)$, $\mu \frac{d}{d \mu}
y_t(\mu) =\beta_{y_t} (y_t,\cdots)$ and $\mu \frac{d}{d \mu}
\lambda(\mu) =\beta_\lambda (\lambda,\cdots)$ for the \MSb couplings
$g_i,y_t$ and $\lambda$ define the \MSb running couplings as functions
of the \MSb energy scale $\mu$. The \MSb renormalization scheme is the
appropriate parametrization for the high energy behavior $\mu \gg M_t$
where $M_t$ is the top-quark mass, the largest of the SM
masses.}. While the gauge couplings behave as expected, $g_1$ as
infrared (IR) free, $g_2$ and $g_3$ as asymptotically (ultraviolet)
free (AF), with leading coefficients exhibiting the related coupling
only, and denoting $c=\frac{1}{16\,\pi^2}$, we have {\small
$$\beta_1=\frac{41}{6}\,g_1^3\,c\simeq 0.00185\semis
\beta_2=-\frac{19}{6}\,g_2^2\,c\simeq -0.00558\semis
\beta_3=-7\,g_3^3\,c\simeq-0.08049\,, $$}
the leading top-quark Yukawa $\beta$-function given by
{\small
\begin{eqnarray*}
  \beta_{y_t}&=&(\frac92\,y_t^3-\frac{17}{12}\,g_1^2\,y_t-\frac94\,g_2^2\,y_t-8\,g_3^2\,y_t)\,c \crn
  &\simeq& 0.02327-0.00103-0.00568-0.07048\crn &\simeq&-0.05391
\end{eqnarray*}
}
\noindent
not only depends on $y_t$ but also on mixed terms with the gauge
couplings which have a negative sign. Interestingly, the QCD
correction is the leading contribution and determines the behavior, to
be opposite from a pure Yukawa behavior. Notice the critical balance
between the dominating strong- and the top-Yukawa couplings: QCD
dominance requires $g_3>\frac{3}{4}\,y_t$ in the gaugeless limit
\mbo{(g_1,g_2=0)}. Similarly, the $\beta$-function of the Higgs
self-coupling, given by {\small
\begin{eqnarray*}
\beta_\lambda&=&(4\,\lambda^2-3\,g_1^2\,\lambda-9\,\lambda\,g_2^2+12\,y_t^2\,\lambda+\frac94\,g_1^4
+\frac92\,g_1^2\,g_2^2+\frac{27}{4}\,g_2^4-36\,y_t^4)\,c \crn &\simeq&
        0.01650-0.00187-0.01961+0.05358+0.00021+0.00149+0.00777\crn &&
        -0.17401 \simeq  -0.11595
\end{eqnarray*}
}
\noindent
is dominated by the top-quark Yukawa contribution and not by the
$\lambda$ coupling itself, again opposite from a pure Higgs system
behavior. Here the sign of the $\beta$-function flips when $\lambda<
\frac32\,(\sqrt{5}-1)\,y_t^2\,$ in the gaugeless \mbo{(g_1,g_2=0)}
limit. This I call true teamwork. It means that running the couplings
up to the Planck scale they have to conspire there such that at long
distances, we can see what we see. This light particle low energy
physics is only possible for specific couplings and by the
self-organized symmetries like local gauge symmetry and chiral
symmetry which let emerge such light states of masses
$M_X\ll\!\!\ll\cdots\MPl$.  We note that all calculations of the
running parameters are based on full 2-loop
matching-conditions~\cite{Yukawa:3,Degrassi:2012ry,Jegerlehner:2012kn,Bednyakov:2015sca,Kniehl:2015nwa}
for the input \MSb couplings and 3-loop RG \MSb
$\beta$-functions~\cite{Mihaila:2012fm,Bednyakov:2012rb,Chetyrkin:2012rz}.
Here one should keep in mind, that the physical parameters extracted
from experimental data and used as an input, in general, do not
include full 2-loop electroweak corrections, so the quoted
uncertainties may well be underestimated and there is definite room
for improvements. For our set of input parameters, the relevant
running \MSb parameters at the Planck scale are of comparable size in
the range 0.51 for $g_2$ being the largest here and 0.35 for $y_t$
being the smallest, with $\sqrt{\lambda}$ at 0.375 slightly larger in
our normalization. This couplings "pattern'' at $\MPl$ is supporting
the view that what was able to penetrate to low energies originates
from a unifying medium.  Note that approximations like the gaugeless
case $(g_1=g_2=0)$ or assuming $\lambda\approx 0$ are not viable
approximations near $\mpl$ neither anywhere at lower energy scales.

\section{Exploiting ``The Infrared - Ultraviolet Connection''}
\label{sec3}
\begin{figure}
\centering
\includegraphics[height=5cm]{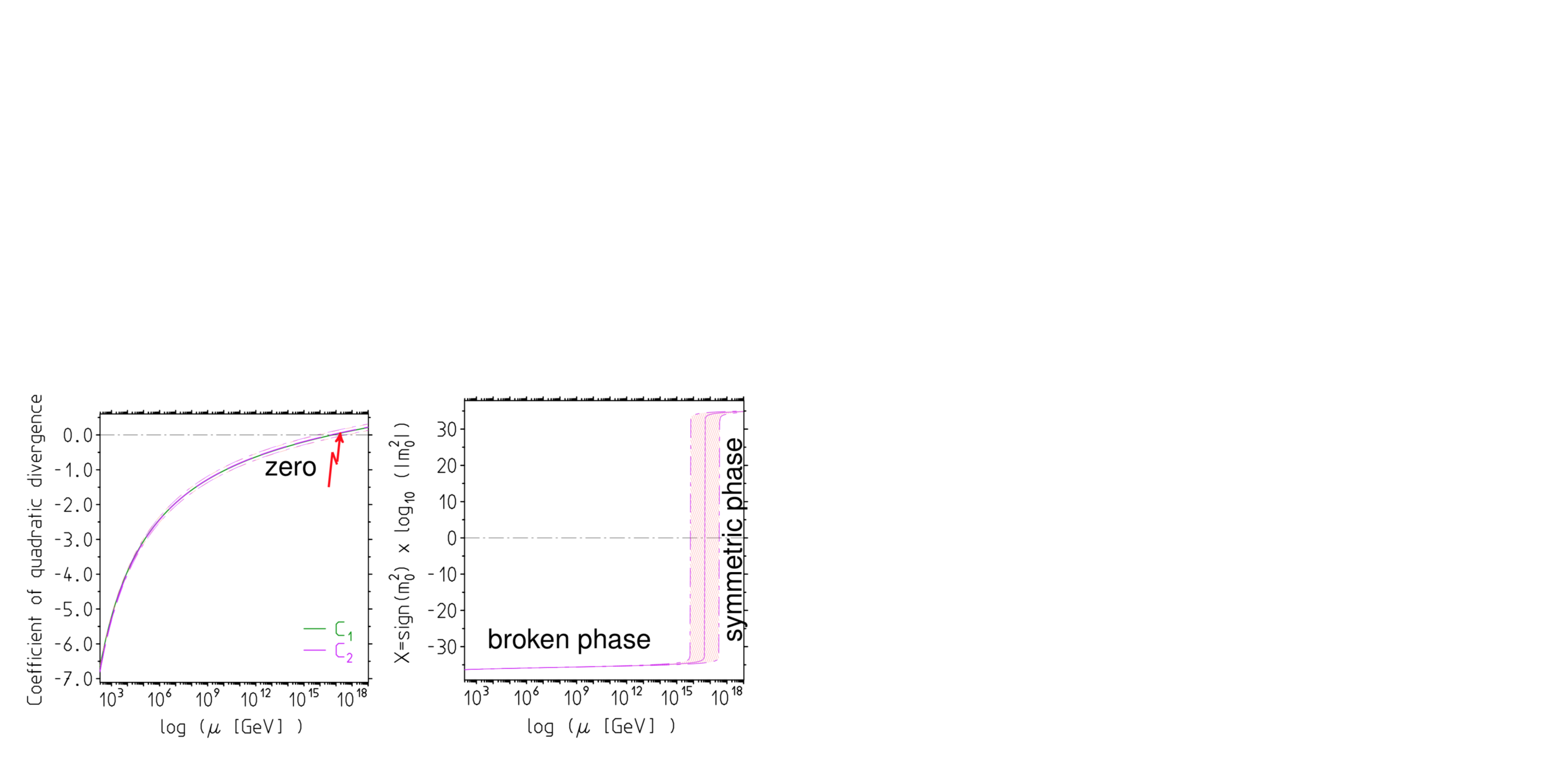}
\caption{The EW phase transition in the SM. Left: the zero in
  \mbo{C_1} and \mbo{C_2} for \mbo{M_H=125.10\pm 0.14\gv}. Right:
  shown is $ X=\sign(m^2_0)\times \log_{10} (|m^2_0|)$. At $\MPl$ we
  have $m_0^2\sim \delta m^2 \simeq
  \frac{\LPl^2}{32\pi^2}\,C(\mu=\LPl) \simeq \left(0.0295\,\lpl
  \right)^2\epo$ A difference between $C_1$ and $C_2$ is barely
  visible on the plot. The shaded bands illustrate the Higgs mass
  dependence in a range [124,126] GeV. Adapted from~\cite{Jegerlehner:2018zxm}.}
\label{Higgsmassjump}
\end{figure}

Veltman's ``The Infrared - Ultraviolet Connection'' concerns the only
quadratic UV divergences of the SM which in the symmetric phase
concerns the mass square term in the Higgs potential \bea
m_0^2=m^2+\delta m^2\;;\;\; \delta m^2= \frac{\LPl^2}{32 \pi^2}\,C
\label{barem2}
\eea which communicates the relationship between the bare $m_0$ (short
distance UV) and the renormalized mass $m$ (low energy IR). The
one-loop coefficient function $C_1$~\cite{Veltman:1980mj} may be
written as (neglecting the small light fermion Yukawa couplings) \bea
C_1={ \frac{6}{v^2}(M_H^2 + M_Z^2 +2 M_W^2-4
  M_t^2)}=2\,\lambda+\frac32\, {g'}^{2}+\frac92\,g^2-12\,y_t^2
\label{coefC1}
\eea and is uniquely determined by dimensionless
couplings. Surprisingly, taking into account the scale dependence of
the SM couplings, the coefficient of the quadratic divergence of the
Higgs mass counterterm exhibits a zero (see Fig.~\ref{Higgsmassjump}
left panel). This has been emphasized
in~\cite{Hamada:2012bp,Jones:2013aua} (see
also~\cite{Wetterich:1983bi}), where the 2-loop result $C_2$ is also
given but represents a minor correction only, i.e. we have
$C(\mu)\simeq C_1(\mu)\simeq C_2(\mu)$. The important point is that in
the LEESM setup the cutoff is physical, meaning that UV singularities
have turned into finite but eventually large numbers.

According to (\ref{barem2}), the SM for the given parameters makes a
prediction for the bare mass parameter $m^2_0=\sign(m^2_0) \times
10^{X}$ of the Higgs potential, where $X$ is displayed in the right
panel of Fig.~\ref{Higgsmassjump}.
In the broken phase we have $ m^2_0=\frac12\,(m^2_{H}+\delta m^2_H)$, which
is calculable! What happens is:
\begin{itemize}[itemsep=1ex, leftmargin=6mm]
\item 
the coefficient \mbo{C(\mu)\simeq C_1(\mu)\simeq C_2(\mu)} exhibits a zero, for $ M_H=125~\gv$ at
about $ \mu_0\sim 1.4 \power{16}~\gv$, clearly but not far below $
\mu=\MPl$,
\item 
at the zero of the coefficient function the counterterm $ \delta
m^2=m^2_0-m^2=0$ ($m$ the \MSb mass) vanishes and the bare mass
changes sign,
\item the sign change \textit{triggers} a \textit{phase transition}
  (PT), the \textit{Higgs mechanism}, that is inducing masses
  simultaneously for the weak gauge-bosons and for all fermions, note
  that the Higgs field exhibits as many different couplings as there
  are different massive SM particles,
\item
  the large \mbo{m_0} and the large $V(0)=\langle V(\phi)\rangle$ (see
  below) for \mbo{\mu > \mu_0} in the symmetric phase initiate
  \textbf{cosmic inflation} when temporarily the potential dominates the kinetic
  term $V(\phi) \gg \frac12\,\dot{\phi}^2$, where $\dot{\phi}=\D \phi/\D t\epo$
\item 
At the transition point $ \mu_0$ we have $m_0=m(\mu_0^2)$ and a bare Higgs
field VEV $v_0=v(\mu_0^2)$, where $ v(\mu^2)$ is the \MSb
renormalized VEV; the power cutoff-effects appear
\underline{nullified} at $\mu_0$! Note that $v$ is characteristic for
long-range order (order parameter) and shows-up for $\mu < \mu_0$
only. In the high energy phase, $\mu >\mu_0$ which extends up to
$\MPl$ we have $v\equiv 0$, no point to expect $v=O(\MPl)$. Where is
the hierarchy problem?
\item
Furthermore there is a jump in the vacuum density, which agrees with
the renormalized one: $-\Delta \rho_{\rm
  vac}=\frac{\lambda(\mu^2_0)}{24}\, v^4(\mu_0^2)\,,$ and thus is
being $O(v^4)$ and \textbf{not} $ O(M^4_{\rm Pl})\,$ as often claimed.
\end{itemize}

\begin{figure}
\centering
\includegraphics[width=0.45\textwidth]{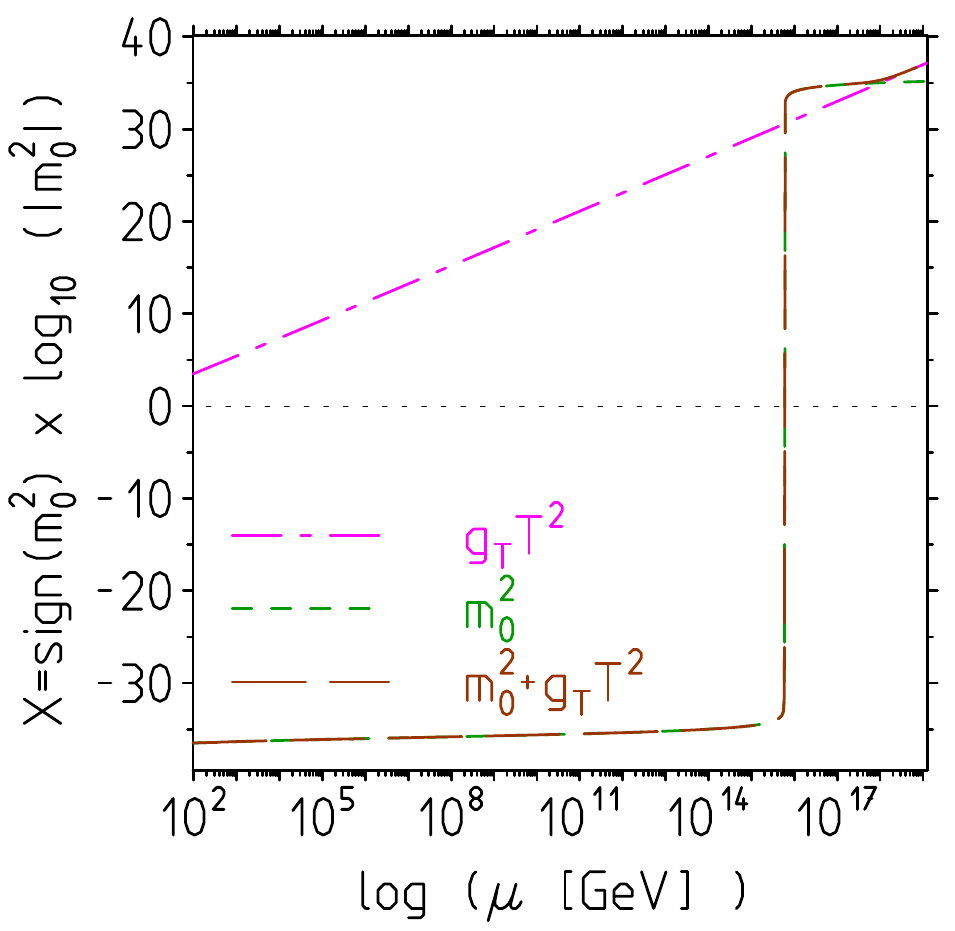}
\includegraphics[width=0.45\textwidth]{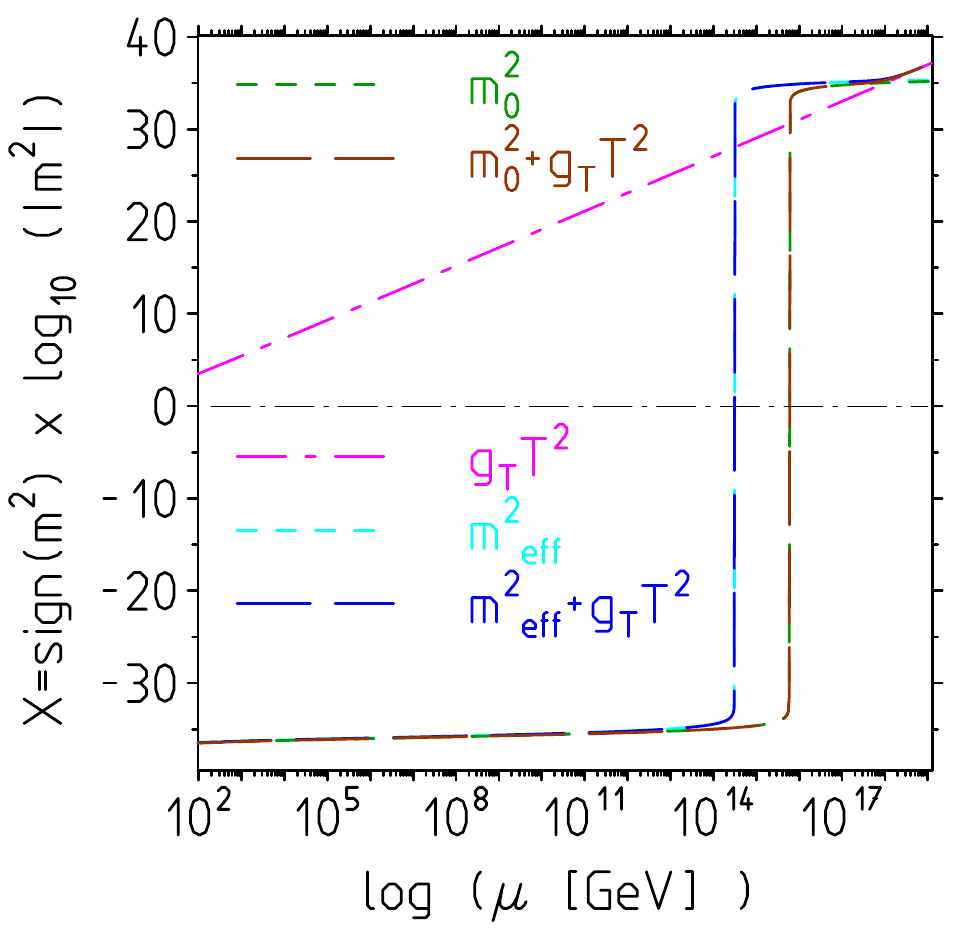}
\caption{\mbo{X} as displayed in the right panel of
  Fig.~\ref{Higgsmassjump} including leading finite temperature
  correction to the potential
  $V(\phi,T)=\ha\,(g_T\,T^2+m_0^2)\,\phi^2+
  \frac{\lambda}{24}\,\phi^4+\cdots$ with $g_T=\frac{1}{16}
  \left[3g_2^2+g_1^2+4y_t^2+\frac23\,\lambda\right]$
  from~\cite{Dine:1992wr} affecting the phase transition point. Left:
  for the bare case [\mbo{m^2,C_1}]. Right: with adjusted effective
  mass from vacuum rearrangement [\mbo{m^2_{\rm eff}={m'}^2,C'_1=C_1+\lambda}]. In
  the case \mbo{\mu_0} sufficiently below \mbo{\mpl}, the case
  displayed here, finite temperature effects affect the position of
  the phase transition little, while the change of the effective mass
  by the vacuum rearrangement is more efficient. The finite
  temperature effect with our parameters is barely visible.  Updated
  from~\cite{Jegerlehner:2018zxm} (where $X$ has been displayed on a
  {\tt log} axis, falsely labled as {\tt log$_{10}$}).}
\label{fig:FT}
\end{figure}

The second important quantity we have not taken into account so far is
the vacuum energy $V(0)=\langle V(\phi)\rangle$ related to the quartic
UV singularity~\cite{Jegerlehner:2014mua}. Again, in the LEESM
scenario, the vacuum energy is a calculable quantity. Considering the
Higgs boson doublet-field $\Phi(x)$, in the symmetric phase $\SU(2)$
symmetry implies that while \mbo{\braket{\Phi(x)}\equiv 0} the
composite field \mbo{\Phi^+\Phi(x)} is a singlet such that the
invariant vacuum energy is represented just by simple Higgs-field
loops, the self-contractions of the Higgs fields in the
potential. With \bea
\bra{0}\Phi^+\Phi\ket{0}=\frac12\bra{0}H^2\ket{0}\equiv
\frac12\,\Xi\semis\Xi =\frac{\lpl^2}{16\pi^2}\,,
\label{vacenergy}
\eea we obtain a cosmological constant (CC) given by
\bea
V(0)=\braket{V(\phi)}=\frac{m^2}{2}\,\Xi+\frac{\lambda}{8}\,\Xi^2=\frac{\lpl^4}{(16\pi^2)^2}\,\frac18 (2C+\lambda)\epo
\eea
A Wick ordering type of rearrangement of the Lagrangian also
leads to a shift of the effective mass
\bea
      {m'}^{2}=m^2+\frac{\lambda}{2}\,\Xi=m^2+ \frac{\LPl^2}{32 \pi^2}\,(C+\lambda)\,\epo
\eea 
For our values of the \MSb input parameters the zero in the Higgs mass
counter term and hence the phase transition point gets shifted
downwards as follows
\bea
{\mu_0 \approx 1.4 \power{16}~\gv} \to
{\mu'_0\approx 7.7 \power {14}~\gv\,.}
\eea
The shift is shown in the right panel of Fig.~\ref{fig:FT} together
with finite temperature effects displayed separately in the left
panel.
We notice that the SM predicts a huge CC at \mbo{\mpl}: \bea
\rho_\phi\simeq V(\phi) \sim (1.29\,\lpl)^4\sim 6.13\power{76}~\gv^4\,,
\eea which is exhibiting a very weak scale dependence (running
couplings) such that we are confronted with the question how to get
rid of this huge quasi-constant? Remember that $\rho_\phi$ has no
direct dependence of $a(t)$ which is the decreasing radius of the
Friedman universe. An intriguing structure again solves the puzzle.
The effective CC counterterm has a zero, which again is a point where
renormalized and bare quantities are in agreement:
\bea
\rho_{\Lambda 0}=\rho_{\Lambda} +\delta \rho_\Lambda \semis \delta \rho_\Lambda
=\frac{\lpl^4}{(16\pi^2)^2}\,X(\mu)
\label{rholambdaCT}
\eea
with $X(\mu)\simeq \frac18\,(2C(\mu)+\lambda(\mu))$ which has a zero
close to the zero of $C(\mu)$ when $2\,C(\mu)=-\lambda(\mu)$. Note
that $C(\mu)=-\lambda(\mu)$ is the shifted Higgs transition point.

Again we find a matching point $\rho_{\Lambda 0}=\rho_{\Lambda}$
between the low-energy and the high-energy world. At this point, the
memory of the quartic Planck scale enhancement gets lost, as it should
be since we know that the low energy phase does not provide access to
cutoff effects.

\begin{figure}
\centering
\includegraphics[width=0.45\textwidth]{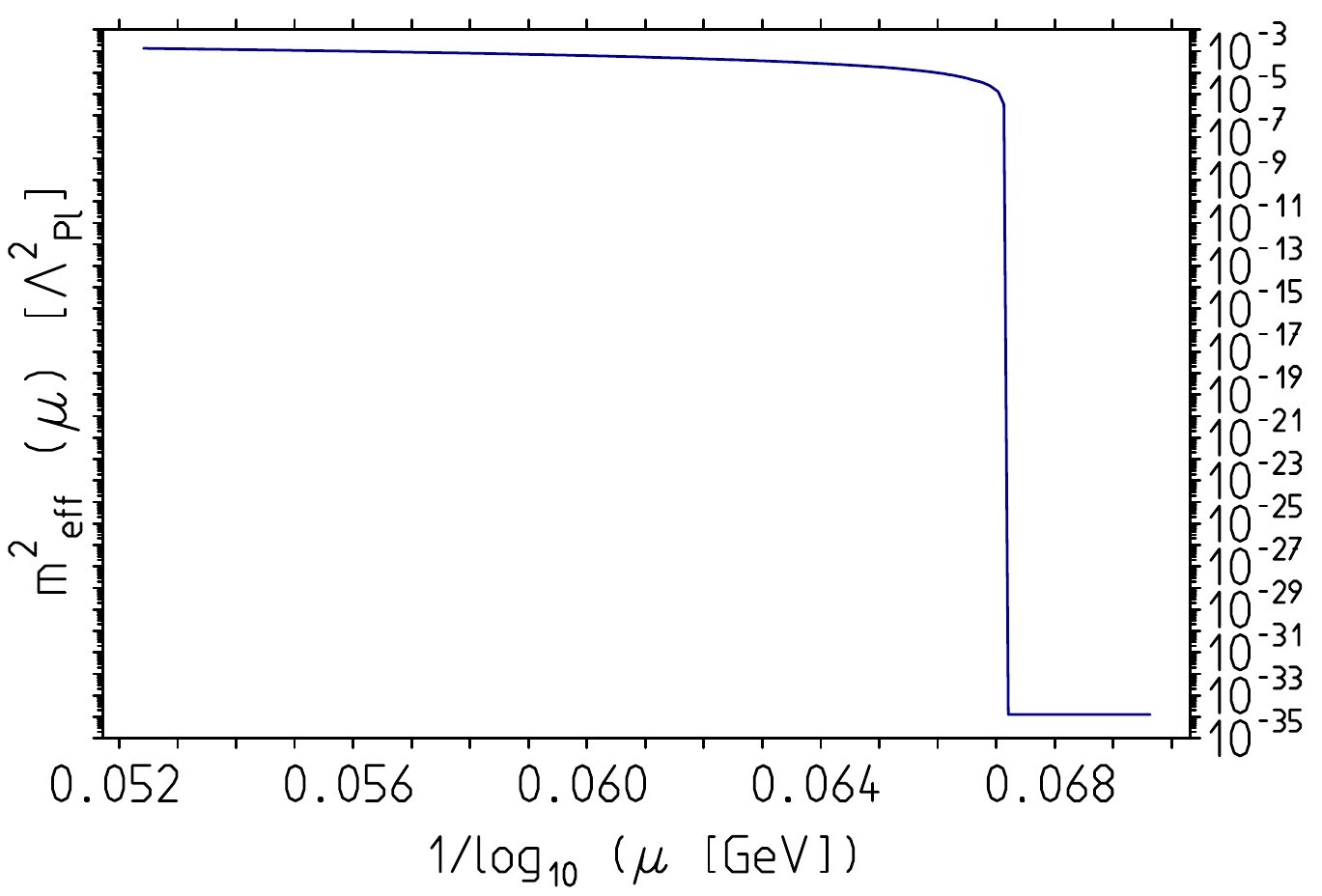}
\includegraphics[width=0.45\textwidth]{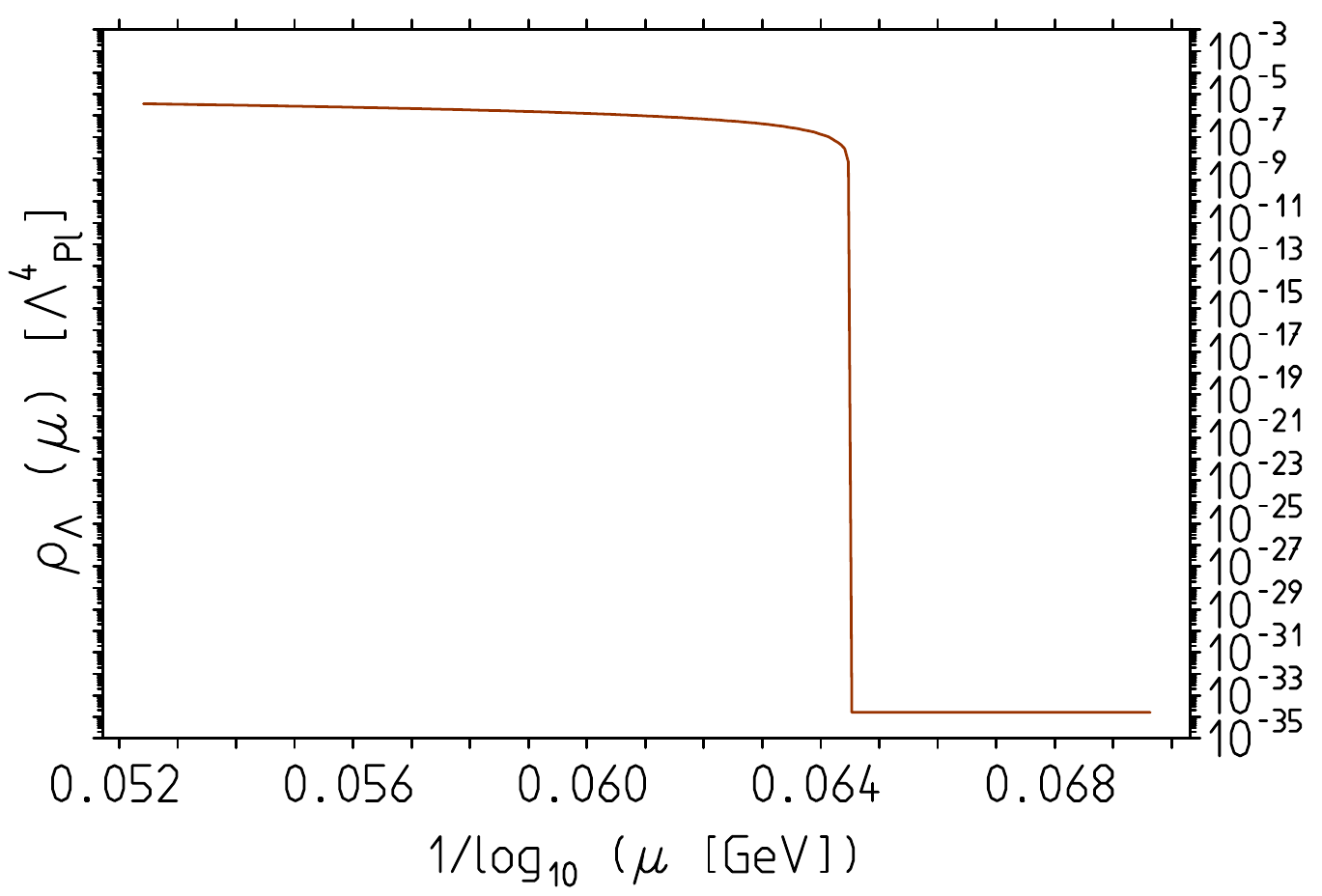}
\caption{The Higgs potential
effective $m^2_{\rm eff}$ [left] and the dark energy density
$\rho_\Lambda$ [right], in units of
\mbo{\lpl}, as functions of ``time'', represented here by $1/\log_{10}\mu\:,$
where $\mu$ represents the energy at that time. Below
the matching point $\mu_{\rm CC}$, where $\rho_\Lambda \simeq 1.6
\power{-47}$ in Planck mass units, we show a scaled up to $\rho_\Lambda
\power{13}$ value of the present dark energy density $\mu_{0\Lambda}^4$
with $\mu_{0\Lambda}\simeq 0.00171~\mathrm{eV}.$
Note: $\rho_\Lambda(t)$ includes besides the large positive
\mbo{V(0)} also negative contributions from vacuum condensates, like
\mbo{\Delta \rho_{\rm EW}} from the Higgs mechanism and \mbo{\Delta
\rho_{\rm QCD}} from the chiral phase transition. Reproduced
from~\cite{Jegerlehner:2018zxm}.}
\label{fig:CCandm2x}
\end{figure}

Crucial point is that
\bea
{X(\mu)=2C+\lambda= 5\,\lambda+3\,g_1^2+9\,g_2^2-24\,y_t^2 }
\label{Xofmu}
\eea acquires positive bosonic contribution and negative fermionic
ones, with different scale-dependence\footnote{Unbroken SUSY would
require a perfect cancellation to happen at all scales. Broken SUSY
would largely eliminate the quadratic and quartic enhancements which
are the driving effects for our scenery. Within the SM the
cancellation between bosonic and fermionic entries only happens at
some point which seems to require fine-tuning. No problem, the energy
scan, the evolution of the universe provides, accomplishes it for
us.}. $X$ can vary dramatically (pass a zero), while individual
couplings are weakly scale-dependent with \mbo{y_t(M_Z)/y_t(\mpl) \sim
  2.7} the biggest and \mbo{g_1(M_Z)/g_1(\mpl) \sim 0.76} the smallest
change. Obviously, the energy dependence of any of the individual
couplings would by far not be able to sufficiently diminish the
originally huge cosmological constant. Only the existence of a zero in
the coefficient function $X(\mu)$ is able to provide the dramatic
reduction of the effective CC, by nullifying the huge cutoff-sensitive
prefactor. Since inflation is tuning the total energy density into the
critical density, the one of a flat geometry, the renormalized dark
energy density can only be a fraction of the critical density, what we
know it is. Fig.~\ref{fig:CCandm2x} illustrates the behavior of the
power enhanced quantities $m^2_{\rm eff}(\mu)=m^2_0(\mu)$ and
$\rho_\Lambda(\mu)$ as a function of ``time'' in units
$1/\log_{10}\mu\,,$ where $\mu$ is the energy scale, decreasing as the
universe is cooling down.

\section{Remark on the impact on inflation}
\label{sec4}
Everybody knows the SM hierarchy problem. True, the SM predicts a
large bare Higgs potential mass in the symmetric phase i.e. for $\mu
>\mu_0$.  Is this a problem? Is this unnatural? In any case, it is a
prediction of the SM! Formally, for $\mu < \mu_0$, in the low energy
phase, (\ref{barem2}) predicts a large negative $m_0^2$ which triggers the
Higgs mechanism, but it has no physical meaning because the physical
Higgs boson mass now is determined by the curvature of the Higgs
potential at its rearranged minimum at $\phi=v$.

At low energy, we see what we see (what is to be seen): the
renormalizable, renormalized SM as it describes close to all we know
up to LHC energies\footnote{Apart from such never clarified 2 $\sigma$
deviations like e.g. the $\sin^2 \Theta_W$ measurement by SLD and LEP,
the new measurement of the muon $g-2$ at Fermilab~\cite{Abi:2021gix}
now reveals a 4.2 $\sigma$ deviation. Similarly, the LHCb
Collaboration observes a 3.1 $\sigma$ deviation in lepton
universality~\cite{Aaij:2021vac}.} and which is devoid of cutoff
effects.  What if we go to very very high energies even close to the
Planck scale? Since for $\mu>\mu_0$ the SM predicts the bare $ m^2_0$
to be positive, we are in the symmetric phase with zero Higgs boson
VEV $v=0$ where we start to see the bare theory i.e. the SM with its
bare short distance effective parameters. In the symmetric phase, all
particle masses vanish except for the now four mass-degenerate
\textbf{very heavy Higgs bosons}. A calculation shows that near below
the Planck scale the potential mass term is dominating and the Higgs
particles can be moving at most very slowly, i.e. the potential
energy\footnote{By $:\!\phi^n\! :$ we denote Wick ordering, which means
that $\phi$-field self-contraction are to be omitted.}
$$V(\phi)=\frac{m^2}{2}\,\phi^2+\frac{\lambda}{24}\,\phi^4
=V(0)+\frac{{m'}^{2}}{2}:\!\phi^2\! :+\frac{\lambda}{24}:\!\phi^4\!
: $$ is large while the kinetic energy $\ha \dot{\phi}^2$ is small.
The Higgs boson contributes to the energy-momentum tensor in
Einstein's gravity equations, providing a pressure
$p=\ha\,\dot{\phi}^2-V(\phi)$ and an energy density $\rho=\ha \,
\dot{\phi}^2+V(\phi)$.  A tick after the Big Bang at Planck time $\tpl$
the slow--roll condition \mbo{\ha \,\dot{\phi}^2\ll V(\phi)} is
satisfied during some time, where $p\approx-V(\phi)\,,\: \rho \approx
+V(\phi)$ so that $p=-\rho$. This means \mbo{\rho=\rho_\Lambda}
represents \textbf{dark energy}! i.e., the system exhibits an unusual
equation of state, which however could be known from ferromagnetic
systems~\cite{Bass:2014lja}. So we learn that the Higgs boson in the
early universe provides huge dark energy (DE)! Note that the latter
includes two different parts $V(\phi)=V(0)+\Delta V(\phi)\,,$ a static
Higgs potential VEV plus a part that depends on the dynamics of the
Higgs field.

The huge DE provides \textbf{anti-gravity} which is inflating the
universe!  Indeed the Friedman equation \mbo{\frac{\D a}{a}=H(t)\,\D
  t} predicts an exponential growth of the radius $ a(t)$ of the
universe as \mbo{a(t)=\exp Ht}, \mbo{H(t)} the Hubble constant
\mbo{H\propto \sqrt{V(\phi)}}. Inflation stops quite quickly as the
field decays exponentially according to the field equation
\mbo{\ddot{\phi}+3H\dot{\phi}\simeq-V'(\phi)}. At some point, the
Higgs potential mass term will be the dominant term
\mbo{V(\phi)\approx \frac{m^2}{2}\,\phi^2} and we have a harmonic
oscillator with friction which predicts Gaussian inflation, as it is
observed in Cosmic Microwave Background (CMB) sky
maps~\cite{Ade:2013uln}. LEESM inflation automatically passes through
the Gaussian phase before inflation halts as it follows from the Higgs
field dynamics. All this tells us that the Higgs boson is the
inflaton, always under the assumption the SM couplings at the end will
turn out to allow for an extrapolation up to $\MPl$.

If inflation happens it tunes the total energy density to be that of a
\textbf{flat space}, which has the particular value \mbo{\rho_{\rm
    crit}=\mu^4_{\rm crit}} with \mbo{\mu_{\rm
    crit}=0.00247\,\mathrm{eV}.} From the CMB data, we know that
today \mbo{\rho_\Lambda=\mu^4_{0,\Lambda}} with
\mbo{\mu_{0,\Lambda}=0.00171\,\mathrm{eV}.} With the ongoing
expansion (cooling) the dark energy density will be approaching
\mbo{\mu_{\infty,\Lambda}=0.00247\,\mathrm{eV}.} In any case, the
large early {cosmological constant} gets tamed by inflation to be
\textit{part of the critical flat space density}. No cosmological
constant problem either? In any case, inflation is proven to have
happened by observation and inflation requires the existence of a
scalar field~\cite{Guth:1980zm,Linde:1981mu}. The Higgs field is precisely such a
field we need and within the SM it has the properties which promote it
to be the inflaton.

All other inflaton models must assume a least one additional scalar
sectors to exist where neither the shape of the potential nor the
parameters are known a priori. In this case, all ``predictions'' have
to be tailored to reproduce what you attempt to predict like e.g. the
spectral indices deriving from CMB data. Higgs inflation is very
different. We know the Higgs boson properties and the mechanism
providing the dark energy. Although this looks pretty straightforward
it is delicate, because of the high sensitivity to the SM parameters
and their uncertainties and missing higher-order corrections. Higgs
inflation is an SM prediction, with all parameters calculable, except
the value of the Higgs field in the Higgs potential. To get the
required amount of inflation, about $N_e\approx 60$ e-folds ($N_e= \ln
[{a(t_{\rm end})}/{a(t_{\rm initial})}]$ with $t_{\rm initial/end}$
the time inflation starts/ends) are required
to cover the CMB causal cone, one needs $\phi_0=4.5\, \MPl$ at Planck
time. This is not unreasonable because the medium is extremely hot
with a Hubble constant of about $H_i\simeq 17\,\MPl$, given by the SM
spectrum and the Stefan-Boltzmann law. The huge field strength at
$\MPl$ decays exponentially in a very short time.

As we know, for arbitrary scales $\mu$, the compensation between
positive bosonic terms and negative fermionic terms within the SM is
incomplete. Exact supersymmetry would amount to a perfect
cancellation in both cases, for $m_0^2$ and $\langle V(0)\rangle$, by
the supersymmetric partners of the SM particles. However, the running
of the SM parameters actually also happens to completely cancel
bosonic and fermionic contributions at some point. This depends on the
proper conspiracy of the SM couplings. That the cancellation only
happens at a particular scale thereby is not important. In fact, the
existence of a matching point solves both the hierarchy problem as
well as the cosmological constant problem.  The nullification of the
power-enhanced terms happens dynamically, without imposing
supersymmetry or alternative mechanisms designed to avoid quadratic
divergences. Such matching points, if they exist below the Planck
scale, are always met at some point since the expansion of the
universe implies the necessary energy scan.

\section{Conclusion}
I have elaborated how Tini Veltman's understanding of the SM structure
provides a convincing picture of the SM as a low energy effective
theory deriving from a Planck medium. Minimality seems to be the
guiding principle and favors the SM as a viable low energy emergent
structure. I do not know any SM extension which is not suffering from
additional fine-tuning issues, while pretended SM naturalness problems
are used to motivate such extensions. Admittedly, also for the SM
parameters, the majority of them particle-Higgs-boson couplings, we
have no idea about what determines their values, which are covering 14
orders of magnitude between the neutrino masses of order $10^{-3}~{\rm
  eV}$ and the top-quark mass at $1.72\power{11}~{\rm eV}\epo$
However, the Higgs boson discovery has told us that there is a
"teamwork" of the leading SM couplings, which can reveal a particular
link of the SM to the Planck world. This may be seen as a hint that a
kind of self-organized system may be at work within the primordial
``ether''\footnote{Let me quote here W{\sc ikipedia}~\cite{wiki}:
Self-organization, also called spontaneous order, is a process where
some form of overall order arises from local interactions between
parts of an initially disordered system. The process can be
spontaneous when sufficient energy is available, not needing control
by any external agent. It is often triggered by seemingly random
fluctuations, amplified by positive feedback. The resulting
organization is wholly decentralized, distributed over all the
components of the system. As such, the organization is typically
robust and able to survive or self-repair substantial
perturbation. Chaos theory discusses self-organization in terms of
islands of predictability in a sea of chaotic
unpredictability.}. Looking at the primordial hot chaotic medium with
the eyes of a distant observer we see the chaos through the filter
which only lets through the long-distance physics. The corresponding
cooperation of the different interactions, balancing bosonic versus
fermionic contributions in running couplings, seems to be even more
striking when we are considering the consequences for the ``Infrared -
Ultraviolet Connection'' first broached by Veltman. The possible
impact of these conspiracies is the SM prediction of Higgs inflation
and reheating after the Big Bang. I have presented only a gross
picture that derives from the leading power-enhanced LEESM
effects. For consideration of finite temperature effects within the
LEESM Higgs-inflation scenario I refer to Fig.~\ref{fig:FT}
and~\cite{Jegerlehner:2013cta,Jegerlehner:2014mua}, which also include
a discussion of reheating and possible impacts for explaining the
baryon-asymmetry in the universe. Also, the bare Higgs potential is
subject to radiative corrections~\cite{Coleman:1973jx} which generates
an effective potential on which the meta-stable vacuum scenarios are
relying. In~\cite{Jegerlehner:2018zxm} I have shown that when
power-enhanced corrections are taken into account in the stable vacuum
case these radiative corrections have little impact on the leading
pattern.

ATLAS and CMS results, the milestone discovery of the SM Higgs boson
together with the absence of any hints of beyond the SM (BSM) physics,
may have unexpectedly revolutionized particle physics, namely
showing that the SM has higher self-consistency (conspiracy) than
expected and previous arguments for the existence of new physics may
turn out not to be compelling. I see the SM as a low energy effective
theory of some cutoff system at $ \MPl$ consolidated.  The crucial
point is the enormous gap between $\MPl$ and from what we can see,
which includes all SM particles. Dark energy and inflation are
unavoidable consequences of the SM provided new physics does not
disturb the SM prediction substantially.

My main theses are:
\begin{itemize}[itemsep=1ex, leftmargin=6mm]
\item[1)] The SM as a renormalized QFT has no hierarchy problem,
  rather its LEESM extension provides a hierarchy solution for
  inflation in the early universe. Over the fairly weak logarithmic
  scale dependence of the running parameters, only the leading
  relevant operators like the Higgs potential and the Higgs boson mass
  term are power enhanced and hence able to ``talk'' with the Planck
  regime conspicuously.
\item[2)] A super-symmetric or any other extension of the SM cannot be
  motivated by the (non-existing) hierarchy problem.
\item[3)] In the early symmetric phase the quadratically enhanced bare
  mass term in the Higgs potential together with the quartically enhanced
  Higgs potential VEV trigger \textbf{inflation} shortly after the Big
  Bang. The relevant time-dependent power corrections are providing a
  strongly time-dependent CC, which is shown in
  Fig.~\ref{fig:CCandm2x}.  This also implies a strongly
  time-dependent Hubble constant in the early universe which possibly
  could solve the existing Hubble constant puzzle~\cite{Poulin:2018cxd}.
\item[4)] As the hot universe cools down after the Big Bang, the
  running of the SM couplings let the bare Higgs potential mass
  $m_0^2$ flip sign from a large positive to a large negative
  value. This is triggering the \textbf{Higgs mechanism} at about
  $\mu_0\sim 10^{16}~\gv$. In the broken phase\footnote{The large
  negative $m_0^2$ ceases to act as a Higgs boson mass and loses its
  observability.  After the memory on the cutoff has been lost at
  that point, within the broken phase the cutoff has no further
  meaning than as a UV regulator and there is no reason to choose it
  to be the Planck cutoff.} the Higgs boson is naturally as light as
  other SM particles which are generated by the Higgs mechanism,
  i.e. including the Higgs boson mass itself all masses are determined
  by the mass-coupling relations $M_W^2=\frac14\,g^2\,v^2,\, M_Z^2=
  \frac14\,(g^2+g'^2)\,v^2,\, M_f^2=\frac12\,y^2_f\,v^2,\,
         \bM_H^2= \frac13\,\blambda\,\bv^2\epo$ Like all other
           particles, the Higgs mass appears generated by the
           non-vanishing VEV \mbo{v \neq 0}. Note that $v$ is
           \textbf{not} Planck scale enhanced as at the Higgs phase
           transition point the renormalized and the bare mass
           coincide (the $\LPl$ enhanced correction is zero at the
           transition point) and in the high energy phase when
           $\mu>\mu_0$ up to $\MPl$ we have $v\equiv 0$.  So, provided
           new physics does not disturb the SM pattern substantially,
           dark energy and inflation are unavoidable consequences of
           the SM Higgs system in our bottom-up approach.
\item[5)] The Higgs mechanism terminates inflation and triggers the
  \textbf{electroweak phase transition}; \textbf{reheating} likely
  proceeds via the four heavy decaying Higgs particles into fermion
  pairs (predominantly top/anti-top quark-pairs) just before the
  system jumps into the broken phase. Again, no non-SM particles
  needed to provide reheating.
\end{itemize}

All that necessitates reconsidering the early pre-Higgs phase epoch of
cosmology. The SM then is in the symmetric phase, which takes a very
different form from known physics in the broken phase. In place of the
massless photon which emerges at the electroweak phase transition, in
the symmetric phase we have all the SM gauge bosons as massless
radiation fields together with the massless Weyl fermions. At these
times there are four very heavy Higgs particles that drive inflation
and reheating.  A more detailed understanding of the EW phase
transition is required and could unveil surprises.

Epilogue: The sharp dependence of the Higgs vacuum stability on the SM
input parameters and possible SM extensions and the vastly different
scenarios which can result as a consequence of minor shifts in
parameter space makes the stable vacuum case a particularly
interesting one and it could reveal the Higgs particle as ``the master
of the universe''. After all, it is commonly accepted that dark energy
is the stuff shaping the universe both at very early as well as at the
late times.

A lot of details have to be worked out to scrutinize the whole
picture. On the SM side, the major issue for the future is the very
delicate \textbf{conspiracy between SM couplings}, which means that
precision determinations of the parameters now turn to be more
important than ever. Precision measurements of top-quark and Higgs
boson properties should be a prime challenge for the LHC and the
ILC/FCC-ee projects. Precision values for \mbo{\lambda}, \mbo{y_t} and
\mbo{\alpha_s} from the high energy frontier should go together with
a program aiming to provide more precise low energy hadronic cross
section measurements at low energy hadron facilities together with
lattice QCD calculations of hadronic vacuum polarization effects which
would allow one to reduce hadronic uncertainties in \mbo{\alpha(M_Z)}
and \mbo{\alpha_2(M_Z)}.

The mayor open issues: where is dark matter? can we explain baryon
asymmetry?  what triggers the see-saw mechanism explaining the smallness
of neutrino masses? what is tuning the strong CP problem?

While these questions could not be answered under the presumptions of
the standard paradigm based on the belief that going to higher
energies reveals more symmetry and a structurally simpler world
exhibiting supersymmetry, higher symmetry groups, super-gravity, and
strings, I think one should reconsider them under the aspect of the
\textbf{paradigm of emergence}, where nature uncovers more complexity at very
high energy and symmetries are emergent at low energies. It is time to
reflect the prejudices which are guiding speculations on BSM physics
on the path towards higher energies and higher precision. Searching
for emergent structures beyond the SM seems to be a more promising
strategy to actually find what is missing in the SM and could be
proven one day. While I argued that the LEESM emergence scenario
conflicts with hierarchy-problem motivated SM extensions, the LEESM
does not exclude BSM physics which we know we have to incorporate in
any case, like \textbf{dark matter} (e.g. in the form of $\SU(4)$ confined
bound states as studied in~\cite{Appelquist:2014jch}), Majorana
neutrinos\footnote{The 3 singlet neutrinos $\nu_R$ needed to allow the
neutrinos to have masses generated by the Higgs mechanism are included
in the SM and must exhibit corresponding Yukawa couplings. The neutral
Weyl-fermions could be Majorana particles and if so they would also
have a singlet Majorana mass term which is allowed by
renormalizability but is not protected by any symmetry. Then like the
Higgs boson mass in the symmetric phase these Majorana particles would
have Planck scale induced very heavy masses which would induce a
see-saw mechanism and explaining the lightweights of the observed
neutrinos in the broken phase~\cite{Schechter:1980gr}.}, axions, etc. Last but not least the
scenario is easy to falsify: find a 4th family fermion, a SUSY
particle, or whatever modifies SM quadratic effects or the SM
parameter running pattern substantially.

Since the SM parameters, as they have been fixed by experimental data
within their uncertainties and to the extent that the conversion to
\MSb parameters theory-wise is under control, in their cooperation are
very~very close (at 1.3 $\sigma$ as inferred
in~\cite{Bednyakov:2015sca}) if not on the spot to match the window of
Higgs vacuum stability and SM extend-ability up to the gravity ruled
Planck scale, the Higgs-inflation option remains an exciting
scenario. Why the relevant parameters, the gauge couplings, the
top-quark Yukawa coupling, and the Higgs potential self-coupling
determined via the Higgs boson mass, should have values barely missing
the option that the scalar field which is required by inflation may be
identified as the SM Higgs boson field remains to be clarified. So
stay tuned, beware of an unstable vacuum. After all, it is the vacuum
on which our cosmos rests!

\noindent
Acknowledgments:\\ I am indebted to Wolfgang Kluge and Oliver B\"ar
for their careful reading of the manuscript and helpful comments. I
also thank Mikhail Kalmykov for many inspiring discussions and
long-time collaboration in electroweak two-loop calculations and in
particular on working out the relationship between on-shell and \MSb
parameters in the SM, which plays a key role in the present article.

\end{document}